\documentclass[aps,prl,twocolumn]{revtex4}
\usepackage{amssymb}

\usepackage{graphicx}

\begin{document}

\title{Fermi surface of CeIn$_3$ above the N\'{e}el critical field}
\author{N.~Harrison$^1$, S.~E.~Sebastian$^2$, C.~H.~Mielke$^1$, A.~Paris$^1$, M.~J.~Gordon$^1$, C.~A.~Swenson$^1$, D.~G.~Rickel$^1$, M.~D.~Pacheco$^1$, P.~F.~Ruminer$^1$, J.~B.~Schillig$^1$, J.~R.~Sims$^1$, A.~H.~Lacerda$^1$, M.-T.~Suzuki$^3$, H.~Harima$^3$, and T.~Ebihara$^4$}
\affiliation{$^1$National High Magnetic Field Laboratory, Los Alamos National Laboratory, MS E536,
Los Alamos, New Mexico 87545\\
$^2$Cavendish Laboratory, University of Cambridge, Madingley Road, Cambridge CB3 0HE, UK\\
$^3$Department of Physics, Kobe University, Kobe 657-8501, Japan\\
$^4$Department of Physics, Shizuoka
University, Shizuoka 422-8529, Japan}
\date{\today}

\begin{abstract}
We report measurements of the de~Haas-van~Alphen effect in CeIn$_3$ in magnetic fields extending to $\approx$~90~T, well above the N\'{e}el critical field of $\mu_0H_{\rm c}\approx$~61~T. The unreconstructed Fermi surface a-sheet is observed in the high magnetic field polarized paramagnetic limit, but with its effective mass and Fermi surface volume strongly reduced in size compared to that observed in the low magnetic field paramagnetic regime under pressure. The spheroidal topology of this sheet provides an ideal realization of the transformation from a `large Fermi surface' accommodating $f$-electrons to a `small Fermi surface' when the $f$-electron moments become polarized. 
\end{abstract}

\pacs{PACS numbers:
..............................}
\maketitle

Strong magnetic fields are an indispensable tool for studying the energy scales relevant to antiferromagnetism. By polarizing their magnetic moments, they deplete the system of available spin degrees of freedom for staggered ordering. In $f$-electron antiferromagnets, polarization is expected to be accompanied by the effective removal of the $f$-electron degrees of freedom from the Fermi surface (FS)~\cite{wasserman1,evans1,edwards1}, enabling access to the underlying electronic structure in its simplest form. The $f$-electron system CeIn$_3$ provides an essential paradigm to understand universal aspects of the relationship between antiferromagnetism and unconventional superconductivity$-$ given the magnetic simplicity of this non-metamagnetic  cubic system~\cite{evans1,ebihara1} compared to CeRu$_2$Si$_2$~\cite{ceru2si2}, CeB$_6$~\cite{goodrich2} or CeRhIn$_5$~\cite{takeuchi1} due to the absence of metamagnetism. Universal behavior is realized in CeIn$_3$ by the application of pressure, whereupon antiferromagnetism is suppressed and superconductivity~\cite{mathur1} and a heavy fermion behavior emerge~\cite{settai1}. If NdB$_6$ provides a model example of a cubic system in which the hybridization between the $f$-electrons and conduction electrons remains negligible throughout~\cite{goodrich1}, then CeIn$_3$ may be considered as the model system for understanding antiferromagnetism preceding superconductivity in the opposite strongly correlated regime comprising heavy quasiparticles~\cite{endo1,sebastian1}.

\begin{figure}[htbp!]
\centering
\includegraphics[width=0.45\textwidth]{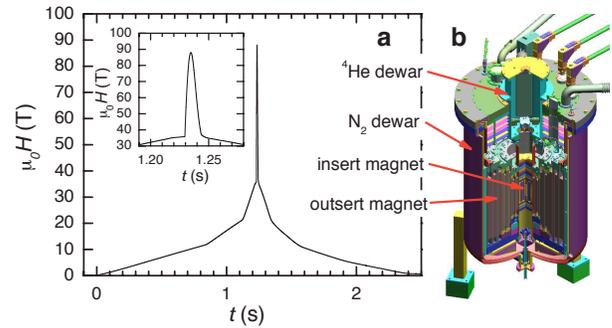}
\caption{{\bf a} The $H$-versus-time $T$ profile of the pulse generated by the combined `outsert' and `insert' magnets. The inset shows the region of the pulse profile provided by the insert magnet in which dHvA measurements in Fig.~\ref{signal01} are made. {\bf b} A schematic of the magnet used for generating the pulse (outer diameter $\approx$~1.4~m).} \label{pulse}
\end{figure}

One unavoidable consequence of a monotonic non-metamagnet magnetization is that much stronger magnetic fields are required to polarize the quasiparticle bands to suppress the correlations~\cite{detwiler1,sakikibara1}. In CeIn$_3$ this requires exceeding the critical field of the N\'{e}el ordered phase, $\mu_0H_{\rm c}\approx$~61~T~\cite{ebihara1}. To determine the electronic structure of CeIn$_3$ in magnetic fields above $H_{\rm c}$, we utilize the recently constructed 100 tesla magnet at Los Alamos~\cite{bacon1}$-$ presently delivering magnetic fields of up to 90~T (see Fig.~\ref{pulse}) while being commissioned. Measurements of the de~Haas-van~Alphen (dHvA) effect over a wide interval in field above $H_{\rm c}$ enable the unreconstructed FS of CeIn$_3$ to be observed in the polarized state, and compared with that previously observed in the paramagnetic regime at pressures exceeding the critical pressure $p_{\rm c}\approx$~26~kbar~\cite{settai1}.

The magnetic field ${\bf H}$ is generated in two stages. First, a 1.4 GW motor-generator is used to energize an `outsert' coil, delivering a $\approx$~36~T `base' magnetic field in a 0.2~m bore. A 2.5~MJ capacitor bank is then used to energize an `insert' coil to produce the remaining $\approx$~54~T in a 15~mm bore~\cite{swenson1}.   Figure~\ref{pulse}a shows an example of the total magnetic field-versus-time profile experienced by the CeIn$_3$ samples studied in this work. With the exception of the magnetic field generation, the dHvA experimental technique is identical to that used in regular pulsed magnetic field experiments~\cite{ebihara1,goodrich1, harrison1}. Three single crystalline CeIn$_3$ samples are cut and etched to diameters of less than 300~$\mu$m for experiments with ${\bf H}\|$$<$100$>$, $<$110$>$ and $<$111$>$. The dHvA effect is measured using a coaxially-arranged compensated pair of detection coils with the innermost coil having $\approx$~460 turns and an inner bore of 450~$\mu$m. A digitizer captures the dHvA signal data while temperatures between 300~mK and 4~K are obtained by controlling the vapor pressure of liquid $^3$He and $^4$He reservoirs. 

\begin{figure}[htbp!]
\centering
\includegraphics[width=0.45\textwidth]{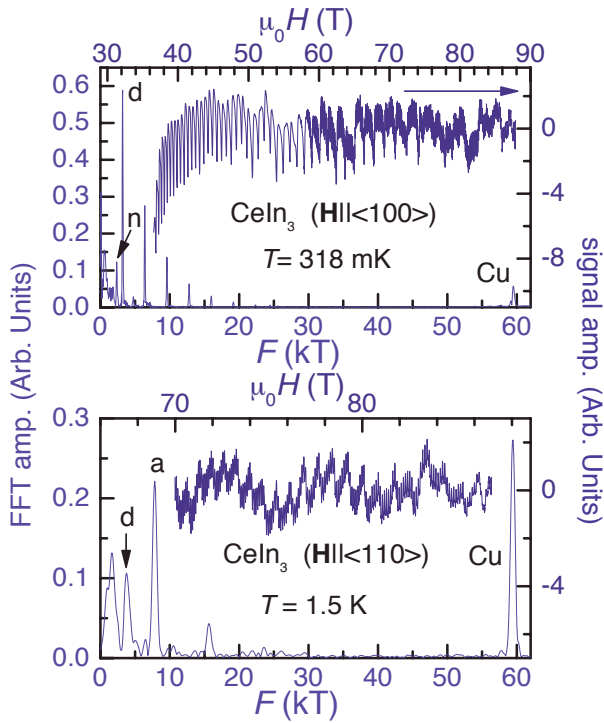}
\caption{Examples of dHvA signal measurements on CeIn$_3$ for two different orientations of ${\bf H}$ together with Fourier transformations. The Cu signal originates from the polycrystalline Cu comprising the detection coils.} \label{signal01}
\end{figure}

\begin{figure}[htbp!]
\centering
\includegraphics[width=0.45\textwidth]{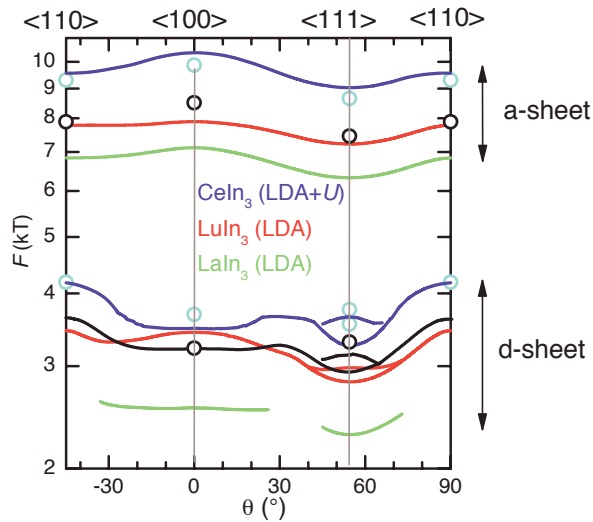}
\caption{A comparison of the d- and a-sheet FS's of CeIn$_3$ measured at ambient pressure and 27~kbar with those calculated for CeIn$_3$ (blue lines), LuIn$_3$ (red lines) and LaIn$_3$ (green lines) using the local density approximation (LDA) method (inclusive of the Coulomb interaction $U$ in the case of CeIn$_3$~\cite{suzuki1}). Black lines indicate the magnetic field-orientation dependence of the d-sheet obtained by Endo {\it et al}~\cite{endo1}, revealing a close similarity to that of LuIn$_3$. Measured frequencies are constant to within 1~\% between 50 and 90~T. Black open circles represent the a- and d-sheets observed by us in strong magnetic fields and ambient pressure, while cyan circles represent the d- and a-sheets measured by Settai {\it et al.} for $p>p_{\rm c}$~\cite{settai1}.}\label{comp01}
\end{figure}

Figure~\ref{signal01} shows examples of dHvA signals and Fourier transforms for ${\bf H}\|$$<$100$>$ and $<$110$>$. For ${\bf H}\|$$<$100$>$, the signal is dominated by the d-branch in both the antiferromagnetic ($\mu_0H<\mu_0H_{\rm c}\approx$~61~T) and polarized paramagnetic regimes ($H>H_{\rm c}$). The n-frequency~\cite{endo1} is also observed to appear prominently at high magnetic fields. The a-sheet (see Fig.~\ref{comp01}) yields a relatively weak feature corresponding to a large electron sheet centered at the R point of the Brillouin zone~\cite{suzuki1}; observable above the level of noise over a restricted interval 75-85~T in magnetic fields (see Fig.~\ref{comp01}).
This frequency becomes more prominent for ${\bf H}\|$$<$110$>$ (Fig. 2, lower panel) and $<$111$>$, appearing at all fields $\mu_0H\gtrsim$~55~T. 

\begin{figure}[htbp!]
\centering
\includegraphics[width=0.45\textwidth]{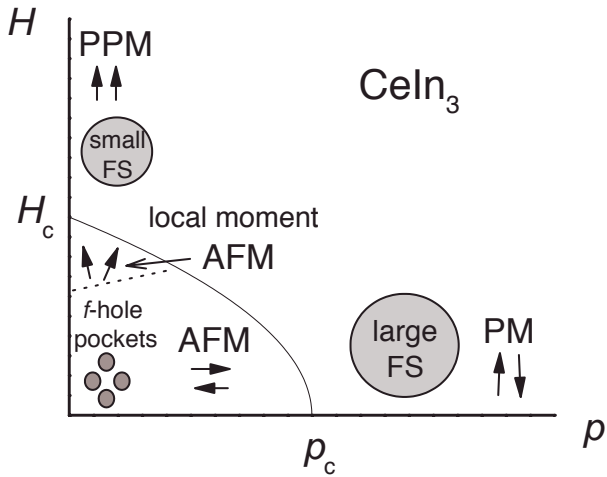}
\caption{A schematic $p$ versus $H$ phase diagram of CeIn$_3$, including the antiferromagnetic (AFM), paramagnetic (PM) and polarized paramagnetic (PPM) regimes. Solid arrows represent the spin states of the $\Gamma_7$ doublet of Ce in each of these regimes, while the grey circles represent the different FS's. The `large FS' includes $f$-electron charge degrees of freedom whereas the `small FS' does not. Small $f$-hole pockets have recently been observed inside the antiferromagnetic phase at ambient pressure~\cite{sebastian1}, but are observed to become depopulated in magnetic fields above $\approx$~41~T (dotted line) where the staggered moment is canted.}\label{diagram01}
\end{figure}

Fermi surface measurements of Ce compounds are often reported to be consistent with either of two dichotomous scenarios. In one scenario, good agreement is found with bandstructure calculations in which the $f$-electron shells are completely filled or empty, as for the Lu and La analog compounds, indicating that the $f$-electrons contribute negligibly to the FS volume. A compound with these characteristics is considered to have a `small FS' (i.e. the FS is much smaller than it might otherwise be were the $f$-electrons to contribute their charge degrees of freedom)~\cite{coleman1}. In the other scenario, some level of agreement is found with bandstructure calculations in which the $f$-electrons are treated as band electrons. A compound with these characteristics is then considered to have a `large FS' (see Fig.~\ref{diagram01})~\cite{coleman1}. Our present measurements outside the antiferromagnetic phase of CeIn$_3$ reveal that both these scenarios are realized in the same isotropic material under conditions of either extreme pressure~\cite{settai1} or intense magnetic field.
dHvA measurements made at $p>p_{\rm c}$ are consistent with band structure calculations in which the $f$-electrons are treated as itinerant, as shown in Fig.~\ref{comp01}~\cite{suzuki1}. Satisfactory agreement requires the effects of Coulomb repulsion and the orbital manifold of the lowest lying $\Gamma_7$ doublet to be taken into consideration~\cite{suzuki1}. Our high magnetic field a-sheet measurements on CeIn$_3$ (see Fig.~\ref{comp01}), in contrast, are found to be similar to the predicted electronic structure of LuIn$_3$, which has filled $f$-shells.  CeIn$_3$ therefore provides a particularly clear example of a system in which a transformation occurs from a `large FS' at high pressures and low magnetic fields to a `small FS' at high magnetic fields and ambient pressure. Since CeIn$_3$ is non-metamagnetic~\cite{evans1,ebihara1} and it is possible (in principle) to move from the high pressure regime to the high magnetic field regime without crossing the antiferromagnetic phase boundary, the transformation in FS must take place in a continuous fashion.

\begin{figure}[htbp!]
\centering
\includegraphics[width=0.45\textwidth]{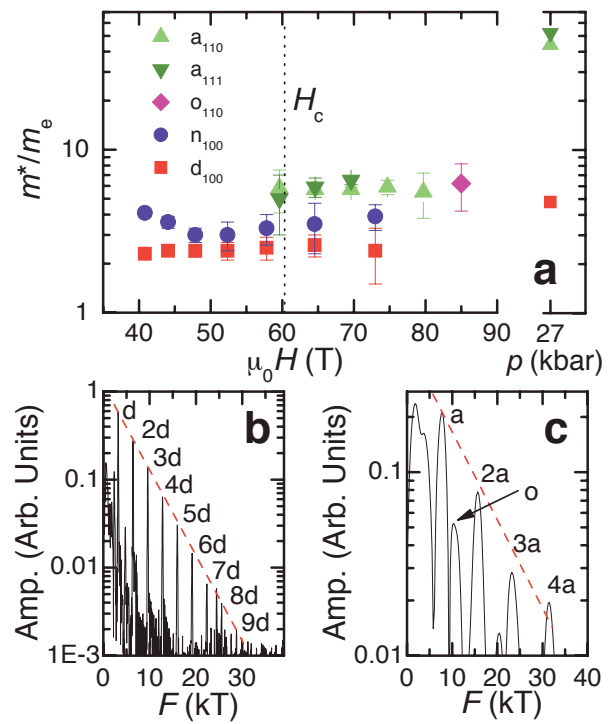}
\caption{{\bf a} Effective masses of different extremal dHvA orbits in CeIn$_3$, estimated by fitting the Lifshitz-Kosevich theoretic form to the temperature dependence of the quantum oscillation amplitude measured between 300~mK and 4~K. The subscript in the legend refers to the orientation of ${\bf H}$. Masses for the same orbits measured at $p>p_{\rm c}$ by Settai {\it et al.} are shown to the right for comparison. {\bf b} Part of the ${\bf H}\|$$<$100$>$ Fourier transform from Fig.~\ref{signal01} plotted on a logarithmic scale so as to show the exponential dependence of the d$_{100}$ frequency harmonics on harmonic index. The red dotted line is a guide to the eye. {\bf c} A similar Fourier transform for the a$_{110}$ frequency, performed over a restricted interval in magnetic field 80-87~T where the harmonic content is most pronounced.}\label{masses01}
\end{figure}

The present experimental limitations require us to study the link between the high pressure and high magnetic field regimes via the intervening antiferromagnetic phase. The manner in which each section of the FS is modified by the antiferromagnetic order parameter depends on its size, location in ${\bf k}$ space and the extent to which it accommodates $f$-electrons. The d-sheet passes through $H_{\rm c}$ and $p_{\rm c}$ in Fig.~\ref{diagram01} relatively unperturbed in topology or effective mass~\cite{settai1,suzuki1} as indicated in Figs.~\ref{signal01} and \ref{masses01}a. 
This robustness to antiferromagnetism and high magnetic fields arises from the minimal contribution of the $f$-electrons to the d-sheet volume (the $f$-electron dispersion exhibits a deep minimum at the $\Gamma$ point in the Brillouin zone~\cite{suzuki1}), and its small size well within the interior of the antiferromagnetic Brillouin zone.

The a-sheet, by contrast, is radically affected by antiferromagnetism owing to its much greater size and hybridization with the $f$-dispersion near the Fermi energy. The large size of the staggered moment within the antiferromagnetic phase of CeIn$_3$~\cite{lawrence1}, combined with the weak dispersion of the $f$-band in the paramagnetic phase~\cite{suzuki1}, requires the antiferromagnetism to be considered from the strong coupling perspective~\cite{sebastian1}. The disappearance of the a-sheet at pressures $p<p_{\rm c}$~\cite{settai1} reflects the effective removal of the majority of the $f$-electrons from the FS deep within the antiferromagnetic phase, where strong coupling gaps the $f$-electron dispersion. Unlike the d-sheet FS, the evolution of the a-sheet FS topology cannot easily be predicted in the intermediate regime close to the antiferromagnetic boundary where local staggered moment ordering competes with Kondo screening~\cite{si1}. A clearer picture begins to emerge in high magnetic fields once the hybridization becomes perturbatively weak due to the field-induced polarization of the quasiparticle bands (as with the suppression of Kondo screening deep within the antiferromagnetic phase~\cite{sebastian1}). The effective mass of the a-sheet in Fig.~\ref{masses01}a is observed to be magnetic field independent (within experimental error) and roughly an order of magnitude smaller than that observed at $p>p_{\rm c}$, providing a compelling evidence for the removal of the majority of the $f$-electrons from this sheet due to the field-induced polarization of the quasiparticle bands~\cite{edwards1}. Whilst the experimental picture at low magnetic fields and ambient pressure is more complex, with small pockets of $f$-holes~\cite{sebastian1} coexisting with fragments of the unhybridized conduction band FS resembling LuIn$_3$, magnetic breakdown tunneling at higher magnetic fields causes the re-emergence of this a-sheet at fields slightly below $H_{\rm c}$.

Spin-dependent effective masses are another consequence of the polarization of the $f$-electrons in strong magnetic fields. In the case of the d$_{100}$ frequency, shown in Fig.~\ref{masses01}b, the absence of a significant $f$-electron contribution causes the spin dependence to closely mimic the localized $f$-electron behavior seen in the single impurity limit, as realized in Ce$_x$La$_{1-x}$B$_6$~\cite{harrison1} and Ce$_x$La$_{1-x}$RhIn$_5$~\cite{alver1} for $x\lesssim$~10~\%. Localization of the $f$-electrons causes the spin-up and -down dHvA frequencies to be the same, but with the lighter mass spin component dominating the dHvA frequency, causing the harmonic index-dependence of the dHvA amplitude to decay in a simple exponential manner. In the case of the a$_{110}$ frequency, four harmonics are observed at the highest magnetic fields, 80~$<\mu_0H<$~87~T, in Fig.~\ref{masses01}c. 
The observed field-independence of the a-sheet FS topology and effective mass suggests that the polarization of the quasiparticle bands is more complete than realized in CeB$_6$ and CeRu$_2$Si$_2$, where  well separated dHvA frequencies corresponding to split spin-up and -down Fermi surfaces and/or field-dependent effective masses are observed~\cite{ceru2si2,endo2}. Exchange splitting effects caused by the polarized $f$-moments (as in NdB$_6$~\cite{goodrich1,gorkov1}) can also not be resolved at high magnetic fields in CeIn$_3$. The new o$_{110}$ frequency (and its harmonic) at $\approx$~10400~T in Fig.~\ref{masses01}c has a similar size to other features predicted in the LuIn$_3$ bandstructure calculations~\cite{harima1}.

In summary, we observe the a-sheet FS of CeIn$_3$ in strong magnetic fields $H>H_{\rm c}$, which is found to be consistent with the `small FS' picture~\cite{coleman1}, in which the $f$-electrons do not contribute significantly to its volume, in contrast to that observed within the paramagnetic regime at pressures $p>p_{\rm c}$.  Consequently, its effective mass is observed to be reduced by an order of magnitude compared to that at $p>p_{\rm c}$. The spheroidal geometry of the FS represents an ideal embodiment of the change in the electronic structure from large FS (at high pressure) to a small FS (in strong magnetic fields). Although a direct observation of this transformation is presently masked by the intervening antiferromagnetic phase, the transformation is expected to take place continuously given the cubic symmetry of CeIn$_3$~\cite{evans1} (the absence of metamagnetism is already established at ambient pressure~\cite{ebihara1}).
The present experiments on CeIn$_3$ show the importance of extreme experimental conditions for understanding electronic structure of strongly correlated $f$-electron metals.

This work was performed under the auspices of the National Science
Foundation, the Department of Energy (US) and Florida state. T.E. acknowledges
support provided by Grant-in-Aid for Scientific Research on priority
Areas, `High Field Spin Science in 100T'  (CASIO) and MEXT. S.E.S.
acknowledges support from the Institute for Complex Adaptive Matter and from
Trinity College, Cambridge University.


\begin{thebibliography}{99}
\bibitem{ceru2si2} Strong anisotropy of the hybridization between the $f$-electrons and the conduction electrons is required to explain metamagnetism in CeRu$_2$Si$_2$ in the absence of magnetic ordering~\cite{evans1}.
\end{thebibliography}
\end{document}